# Electrons with Planckian scattering obey standard orbital motion in a magnetic field

Amirreza Ataei[1], A. Gourgout[1], G. Grissonnanche[1], L. Chen[1], J. Baglo[1], M-E. Boulanger[1], F. Laliberté[1], S. Badoux[1], N. Doiron-Leyraud[1], V. Oliviero[2], S. Benhabib[2], D. Vignolles[2], J.-S. Zhou[3], S. Ono[4], H. Takagi[5,6,7], C. Proust[2], Louis Taillefer[1,8]

*1 Institut quantique, Département de physique & RQMP, Université de Sherbrooke, Sherbrooke, Québec, Canada*

*2 LNCMI-EMFL, CNRS UPR3228, Univ. Grenoble Alpes, Univ. Toulouse, INSA-T, Grenoble and Toulouse, France*

*3 Texas Materials Institute, University of Texas - Austin, Austin, Texas, USA*

*4 Central Research Institute of Electric Power Industry (CRIEPI), Nagasaka, Yokosuka, Japan*

*5 Max Planck Institute for Solid State Research, Stuttgart, Germany*

*6 Department of Physics, University of Tokyo, Tokyo, Japan*

*7 Institute for Functional Matter and Quantum Technologies, University of Stuttgart, Stuttgart, Germany*

*8 Canadian Institute for Advanced Research, Toronto, Ontario, Canada*

In various "strange" metals, electrons undergo Planckian dissipation [1,2], a strong and anomalous scattering that grows linearly with temperature [3], in contrast to the quadratic temperature dependence expected from the standard theory of metals. In some cuprates [4,5] and pnictides [6], a linear dependence of the resistivity on magnetic field has also been considered anomalous – possibly an additional facet of Planckian dissipation.

Here we show that the resistivity of the cuprate strange metals $Nd_{0.4}La_{1.6-x}Sr_xCuO_4$ [7] and $La_{2-x}Sr_xCuO_4$ [8] is quantitatively consistent with the standard Boltzmann theory of electron motion in a magnetic field, in all aspects – field strength, field direction, temperature, and disorder level. The linear field dependence is found to be simply the consequence of scattering rate anisotropy. We conclude that Planckian dissipation is anomalous in its temperature dependence but not in its field dependence. The scattering rate in these cuprates does not depend on field, which means their Planckian dissipation is robust against fields up to at least 85 T.

The hallmark of "strange metals" is a perfectly linear temperature dependence of the electrical resistivity as temperature goes to zero, in sharp contrast with the $T^2$ dependence expected from the standard Fermi-liquid theory of metals. This phenomenon is called Planckian dissipation because in all cases an estimate [1,2], or a measurement [3], of the inelastic scattering time $\tau$ yields $\tau \sim \hbar / k_B T$, where $\hbar$ is Planck's constant and $k_B$ is Boltzmann's constant. The microscopic mechanism that underlies Planckian dissipation is still unknown, but the beautiful simplicity and universal character of the phenomenon point to a fundamental quantum principle.

It has been suggested that the dependence of resistivity on magnetic field $B$ is another facet of Planckian dissipation in strange metals. Specifically, the scattering rate would have not only an anomalous $T$-linear dependence, but also an anomalous $B$-linear dependence. This suggestion was inspired by the observation of $B$-linear resistivity in cuprates such as $La_{2-x}Sr_xCuO_4$ (LSCO) [4], $Tl_2Ba_2CuO_{6+\delta}$ (Tl2201) and $Bi_2Sr_2CuO_{6+\delta}$ (Bi2201) [5], and pnictides such as $BaFe_2(As_{1-x}P_x)_2$ [6] – a behaviour that contrasts with the usual $B^2$ dependence observed in simple metals. In one proposal, $T$ and $B$ would be linked via a scattering rate of the form $\sqrt{(\alpha k_B T)^2 + (\gamma \mu_B B)^2}$, where $\mu_B$ is the Bohr magneton, and $\alpha$ and $\gamma$ are coefficients of comparable magnitude [6].

To determine whether the linear magnetoresistance is anomalous, we must compare it to what is expected from the standard Boltzmann theory of electron motion in a magnetic field providing all electronic parameters are known. Here we carry out such a comparison in detail for two closely related strange metals: the cuprates $La_{1.6-x}Nd_{0.4}Sr_xCuO_4$ (Nd-LSCO) and LSCO, at a hole concentration (doping) $p = 0.24$.

At that doping, Nd-LSCO is in its purely metallic phase, without pseudogap [9,10,11], charge density wave modulations [12,13] or static magnetism [14] (Fig. 1a). Its superconductivity can be entirely suppressed by applying a magnetic field in excess of 20 T.

Its resistivity is perfectly $T$-linear, down to the lowest temperature ($T \sim 1$ K) [7]. Thermal conductivity measurements down to 50 mK have shown this $T$-linearity to persist down to $T = 0$ [15]. Nd-LSCO is an archetypal strange metal, with a simple quasi-2D single-band Fermi surface, mapped out by ARPES measurements [10,16].

A first test of Boltzmann theory was recently carried out on Nd-LSCO at fixed field, by measuring its angle-dependent magnetoresistance (ADMR) [3]. Changes in the $c$-axis resistivity $\rho_c$ as a function of the field angle relative to the $c$ axis ($\theta$) and $a$ axis ($\phi$) were used to extract the detailed shape and size of the Fermi surface, using the standard Chambers formalism. The resulting Fermi surface was in good agreement with that seen by ARPES, thereby validating Boltzmann theory at fixed field.

In addition, the ADMR data were also used to extract the scattering rate $1/\tau$. Its $T$ dependence was found to be linear, with a Planckian slope, namely $1/\tau = \alpha\, k_B T/\hbar$ with $\alpha \sim 1$ (specifically, $\alpha = 1.2 \pm 0.4$) [3]. Moreover, and crucially, this $T$-linear inelastic scattering rate was found to be isotropic (independent of $\phi$), thereby explaining how a perfect $T$-linear resistivity is possible in a metal whose Fermi surface, density of states and Fermi velocity are strongly anisotropic. The scattering rate is the sum of an elastic ($T$-independent) term and an inelastic ($T$-dependent) term:

$$\frac{1}{\tau(\phi,T)} = c\left[\frac{1}{\tau_0} + \frac{1}{\tau_{aniso}}|\cos(2\phi)|^\nu\right] + \alpha\, k_B T/\hbar \ . \quad (1)$$

Fits to the ADMR data in Nd-LSCO $p = 0.24$ [3] yielded the parameters given in Extended Data Table I, with $c = 1.0$ (by definition) and $\alpha = 1.2$.

The strongly anisotropic elastic term in Nd-LSCO is attributed to the nearby van Hove singularity (together with small-angle scattering [17]), which causes the angle-dependent density of states to be strongly anisotropic, with a maximum in the antinodal directions [3].

Given the Fermi surface and the scattering rate, we can now use Boltzmann theory to predict how the resistivity of Nd-LSCO should vary as a function of field strength, disorder level and field direction, at various temperatures. As shown below, we will find that all predictions are precisely confirmed by our data on Nd-LSCO and on the closely related material LSCO. In other words, the behaviour of electrons in a magnetic field in these strange metals is entirely the result of their orbital motion, and there is no evidence that the scattering rate has any field dependence.

The in-plane resistivity of the three samples considered here is displayed in Fig. 1b. It is perfectly $T$-linear below 70 K in all cases, with similar slopes. The only difference is the residual resistivity (at $T = 0$), which reflects the different levels of disorder (elastic scattering): $\rho_0 = 28, 12$ and $48 \, \mu\Omega$ cm for Nd-LSCO, LSCO sample S1 and LSCO sample S2, respectively.

In Fig. 2a, we display the field dependence of the in-plane resistivity $\rho$ for Nd-LSCO, plotted as $\rho(B)/\rho(0)$, the relative magnetoresistance (MR), obtained by applying a pulsed field up to 85 T, at various constant temperatures (for the full set of isotherms, see Extended Data Figure 1). In Fig. 2b, we show the corresponding prediction of Boltzmann theory, based on the parameters established by ADMR in Nd-LSCO. We see that data and calculation are in excellent quantitative agreement: the MR at 4 K and 80 T is $\rho(B)/\rho(0) = 1.35$ and $1.30$, respectively. Qualitatively, we find that the MR increases with decreasing $T$, and it evolves from $B^2$ at high $T$ to $B$-linear at low $T$ – an evolution that is nicely reproduced by the calculation.

The $B$-linear dependence at low $T$, hailed as anomalous in previous studies, is in fact entirely accounted for by Boltzmann theory, given the strongly anisotropic elastic scattering rate of Nd-LSCO. Indeed, if we remove the anisotropic part of the scattering (by setting $1/\tau_{aniso} = 0$ in Eq. 1), then we lose the $B$-linear character of the MR (Fig. 2d). The fact that the MR becomes quadratic at high $T$ (MR ~ $B^2$ at 100 K) is also accounted for by the calculation (Fig. 2b), and this is due to the loss of anisotropy as the *isotropic* inelastic scattering dominates more and more with increasing temperature. We conclude that in overdoped Nd-LSCO and LSCO, there is no need for $1/\tau$ to depend on field to explain quantitatively the $B$-linearity since it is simply due to the orbital motion of electrons in the presence of anisotropic impurity scattering. In other words, Planckian dissipation in these cuprates is insensitive to field, up to at least 85 T.

In order to directly compare with prior work on LSCO [4], we have also measured the field dependence of $\rho$ in LSCO at $p = 0.24$, in our two samples S1 and S2. In those two samples, $\rho$ is $T$-linear below ~ 70 K, exactly as in Nd-LSCO, with a very similar slope (Fig. 1b). In a prior study on LSCO at $p = 0.23$ [8], in which a field of 48 T was applied to suppress superconductivity, the $T$-linear dependence of $\rho$ was found to extend down to at least $T$ ~ 1 K. Note that the Fermi surface of LSCO is very similar to that of Nd-LSCO [16], namely it is electron-like, since the Fermi level has crossed the van Hove singularity.

In Fig. 2c, we display our high-field data on LSCO S1. The behavior of the MR at various temperatures is seen to be very similar to that found in Nd-LSCO and in the calculations, namely $B$-linear at 4 K evolving to $B^2$ at 100 K. (Note that our MR data on LSCO are also consistent with prior MR data on LSCO $p = 0.19$ [4]; see Extended Data Figure 2.) The only difference is the magnitude of the MR, equal to 1.65 in LSCO S1 vs 1.3 in Nd-LSCO, at 4 K and 60 T. This quantitative difference is expected, given the lower $\rho_0$ in the former sample.

In Fig. 3a, we compare the MR in our three samples, at $T = 30$ K. In Fig. 3b, we show the predicted dependence of the MR on the disorder level. The calculation is done using all the same ADMR-determined parameters, but now varying the multiplicative factor $c$ in front of the elastic term in Eq. 1. By definition, $c = 1.0$ is the value determined by the ADMR study for an Nd-LSCO sample with a very similar $\rho_0$ value to our own Nd-LSCO sample (from the same source and batch). We see that by decreasing the strength of disorder scattering by a factor 0.43, namely the ratio of $\rho_0$ values in our Nd-LSCO and LSCO S1 samples (*i.e.* 12 / 28), so setting $c = 0.43$ in Eq. 1, the calculation reproduces perfectly the MR measured in sample S1. Similarly, setting $c = 1.72$ (*i.e.* 48 / 28), yields a calculated curve in excellent agreement with the MR measured in sample S2. We conclude that Boltzmann theory is able to account very well for the effect of disorder on the magnitude of the MR in these strange metals.

If the MR in Nd-LSCO and LSCO is entirely the result of the orbital motion of electrons around the Fermi surface, this orbital MR should all but vanish when the magnetic field is applied parallel to the $CuO_2$ planes of the layered cuprate structure. If the materials were truly two-dimensional, no motion could arise perpendicular to the planes and so no orbital motion could be induced by a field $B \parallel a$. These cuprates are in fact quasi-2D materials and their Fermi surface has some warping along the $c$ axis, due to a small but nonzero dispersion perpendicular to the planes, such that $\rho_c / \rho_a \approx 250$ (at $p = 0.24$) [7]. In Fig. 4a, the MR predicted for Nd-LSCO is shown for $B \parallel a$ at $T = 50$ K. A huge difference in MR between $B \parallel c$ and $B \parallel a$ is obtained, the latter being smaller by a factor ~ 300 (at $B = 50$ T). In Fig. 4b , we display our high-field data on LSCO S1, for $B \parallel c$ and $B \parallel a$. We observe a huge difference between the two field directions, with a negligible MR when $B \parallel a$. We conclude that all the field dependence of the electrical resistivity in these strange metals is coming from the orbital motion of electrons.

In this respect, the prior report of a large MR for $B \parallel a$ in the cuprates Tl2201 and Bi2201 [5] – comparable to the MR for $B \parallel c$ – is in striking contrast to our data and remains to be understood. Note that Boltzmann calculations for Tl2201 based on ADMR-determined parameters yield a predicted magnetoresistance that is not consistent with the measured magnetoresistance [5].

In summary, the standard Boltzmann theory accounts in detail and quantitatively for all aspects of the magnetoresistance in two archetypal cuprate strange metals, including the dependence on field angle ($\theta$ and $\phi$), on field strength and on disorder level, for temperatures down to $T \sim 0$. While the microscopic mechanism responsible for the perfect $T$-linear dependence of the resistivity in strange metals is still unknown, the strongly interacting electrons that undergo Planckian scattering nevertheless conform to the standard orbital motion in a field, as prescribed by their Fermi surface, Fermi velocity and scattering rate, the latter being independent of field. Planckian dissipation in cuprates is insensitive to magnetic field, at least up to 85 T.

This insensitivity to field sheds new light on the nature of scattering in cuprates. It is natural to associate the scattering process at $p = 0.24$ with the critical doping at which the pseudogap phase ends in Nd-LSCO, namely $p^* = 0.23$ [9]. This endpoint displays the standard thermodynamic signatures of a quantum critical point, with a sharp peak in the specific heat $C$ vs $p$ at $p^*$ and a $\log T$ dependence of $C/T$ at $p^*$ [18]. It was shown that at $p^*$, $C$ is independent of magnetic field up to 18 T [18]. We now find that $\tau$ is independent of field up to 85 T. This is dramatically different from the quantum criticality of other strange metals, like the heavy-fermion metals CeCu$_6$ and CeCoIn$_5$, where a small magnetic field strongly perturbs both the resistivity and the specific heat [19,20]. Clearly, the fluctuations associated with the pseudogap critical point in hole-doped cuprates, presumably responsible for the inelastic $T$-linear scattering near $p^*$ and potentially involved in the $d$-wave pairing [21], are remarkably robust against magnetic fields.

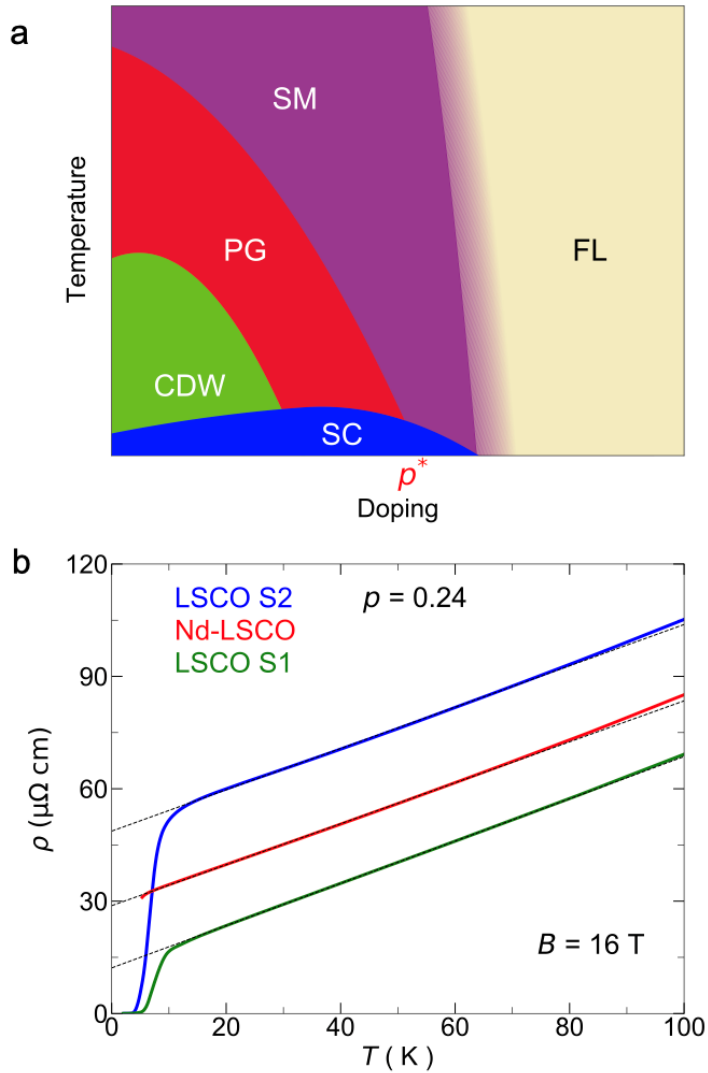

**Fig. 1 | Cuprate phase diagram and *T*-linear resistivity in Nd-LSCO and LSCO.**

**a)** Schematic temperature-doping phase diagram of the cuprate Nd-LSCO, showing the pseudogap phase (PG) [11], the superconducting phase in zero field (SC), the charge-density-wave region (CDW) [12,13], and roughly the region of strange metal behaviour (SM), distinct from the Fermi-liquid behavior (FL). **b)** Temperature dependence of the in-plane resistivity $\rho$ ($J \parallel a$) in a magnetic field $B = 16$ T normal to the copper oxide planes ($B \parallel c$), for our three cuprate samples, all with a doping $p = 0.24$: Nd-LSCO (red); LSCO S1 (green); LSCO S2 (blue). All three exhibit a perfect *T*-linear dependence below $T \sim 70$ K, with a very similar slope. The residual resistivity extrapolated from a linear fit in the interval 20-70 K (dotted lines) is $\rho_0 = 28$, 12 and 48 $\mu\Omega$ cm, respectively. The drop in $\rho$ to zero below 10 K is due to superconductivity, not entirely suppressed at this relatively low field.

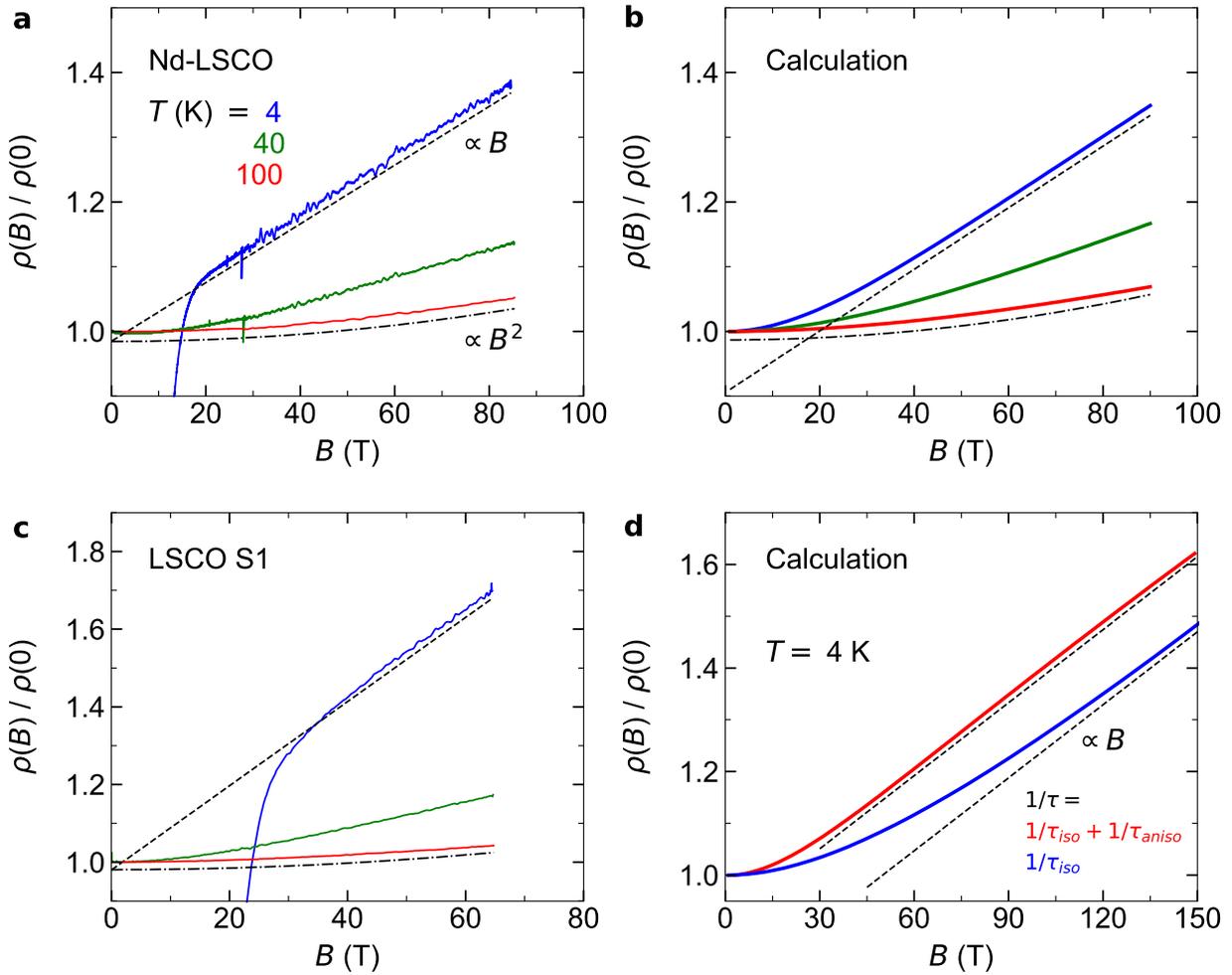

**Fig. 2 | Field dependence of resistivity at various temperatures.**

Measured and calculated magnetoresistance (MR), plotted as $\rho(B)/\rho(0)$ vs $B$, for $J \parallel a$ and $B \parallel c$, at various fixed temperatures, as indicated. **a)** Isotherms measured in Nd-LSCO up to 85 T, for $T = 4$ K (blue), 40 K (green) and 100 K (red). The MR at 4 K is seen to be linear in field above ~40 T, whereas the MR at 100 K is quadratic, as emphasized by the linear (dashed) and quadratic (dashed dotted) lines. **b)** Calculated MR using the parameters for Nd-LSCO extracted from a prior ADMR study [3], for the same three temperatures. **c)** Isotherms measured in LSCO S1 up to 65 T, for the same three temperatures as in panel a. **d)** Calculated MR at $T = 4$ K using the full scattering rate of Nd-LSCO, given in Eq. 1 (red, same as in panel a), and using only the isotropic part of that scattering rate (blue). The parallel dashed lines are linear, to emphasize the loss of $B$-linearity in the isotropic case.

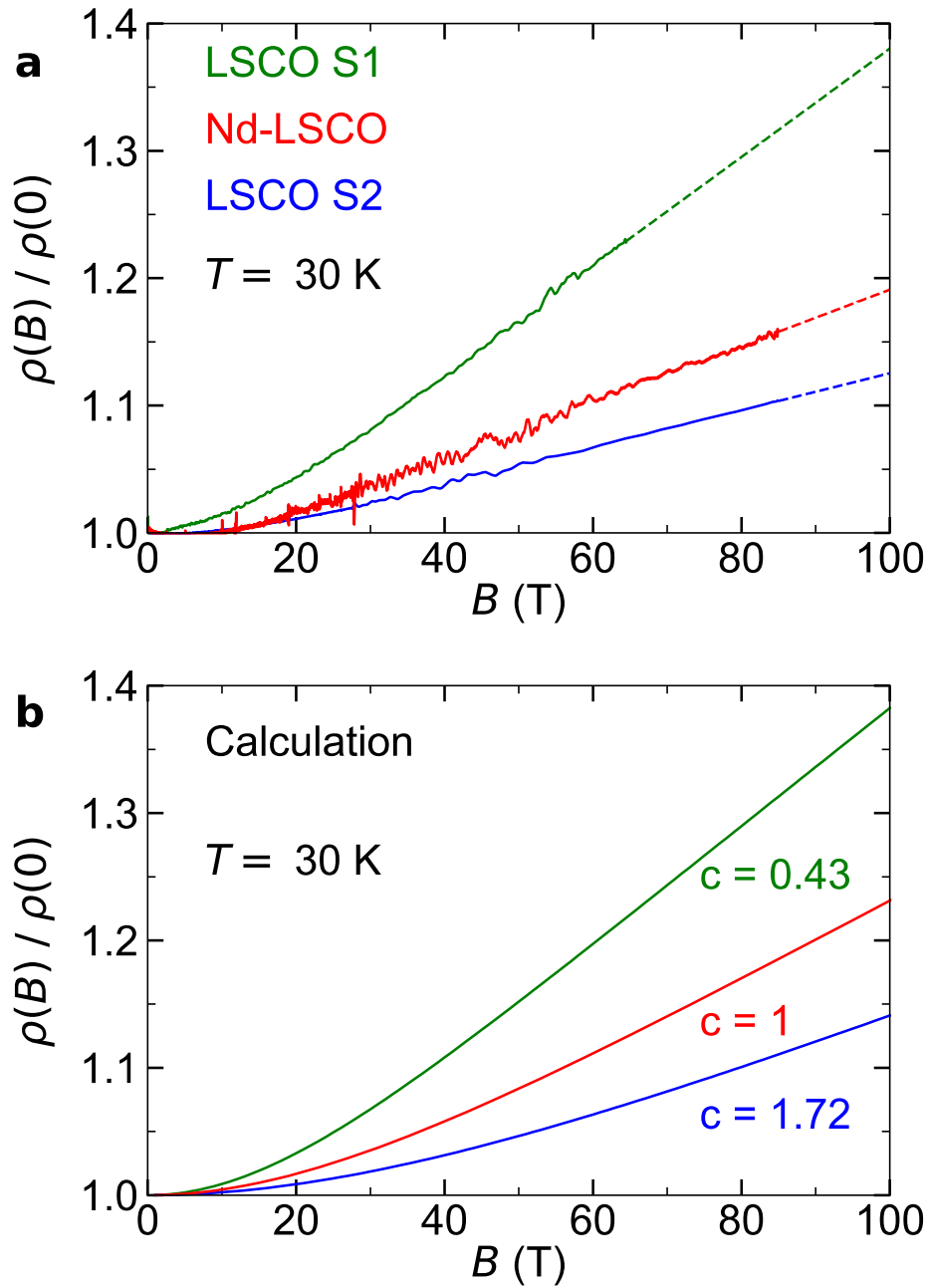

**Fig. 3 | Effect of disorder.**

**a)** Measured magnetoresistance (MR) at $T = 30$ K, for Nd-LSCO (red), LSCO S1 (green) and LSCO S2 (blue). The dashed lines are linear extensions of the data. **b)** Calculated MR at $T = 30$ K, for three levels of disorder, obtained by setting the prefactor of the elastic scattering term in Eq. 1 to $c = 1.0$ (red), 0.43 (green) and 1.72 (blue). The values of $c$ correspond to the variation in the measured $\rho_0$ values of our three samples (Fig. 1b).

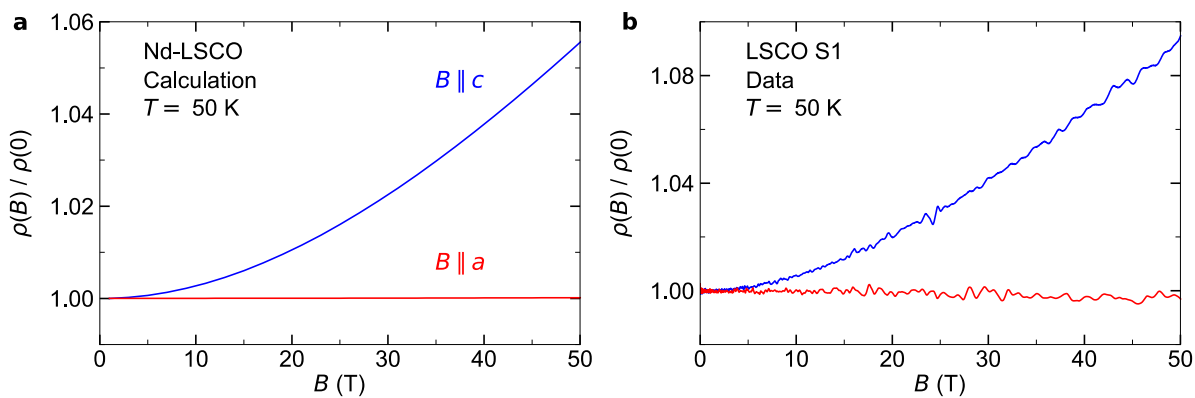

**Fig. 4 | In-plane magnetic field.**

**a)** Calculated normalized MR at $T = 50$ K, for a magnetic field applied parallel (red; $B \parallel a$) and perpendicular (blue, $B \parallel c$) to the CuO$_2$ planes. **b)** Measured magnetoresistance, $\rho(B)/\rho(0)$, at $T = 50$ K, for a magnetic field applied parallel (red; $B \parallel a$) and perpendicular (blue, $B \parallel c$) to the CuO$_2$ planes in LSCO sample S1 at $p = 0.24$. No MR is detected in this sample up to 50 T for $B \parallel J \parallel a$.

**Acknowledgements** A portion of this work was performed at the LNCMI, a member of the European Magnetic Field Laboratory (EMFL). C.P. acknowledges funding from the French ANR SUPERFIELD, and the LABEX NEXT. L.T. acknowledges support from the Canadian Institute for Advanced Research (CIFAR) as a Fellow and funding from the Natural Sciences and Engineering Research Council of Canada (NSERC; PIN:123817), the Fonds de recherche du Québec - Nature et Technologies (FRQNT), the Canada Foundation for Innovation (CFI) and a Canada Research Chair. This research was undertaken thanks in part to funding from the Canada First Research Excellence Fund.

**Author Information** The authors declare no competing financial interest. Correspondence and request for materials should be addressed to A.A. (amirreza.ataei@USherbrooke.ca) or L.T. (louis.taillefer@USherbrooke.ca).

**Data availability.** The data and the program codes that support the experimental data and simulations in the paper are available from the corresponding authors upon a reasonable request.

**Author Contributions** A.A., A.G., S.B., J.B., L.C., ME.B., V.O., D.V. and C.P. performed the high field transport measurements at LNCMI, Toulouse. A.A., A.G., F.L., S.B. performed the transport and characterization measurements at Sherbrooke. J-S.Z. prepared the Nd-LSCO sample, S.O. prepared the LSCO S1 sample, H.T. prepared the LSCO S2 sample. A.A. analyzed the data and made the simulation figures in consultation with G.G. and L.T. A.A. and L.T. wrote the manuscript, in consultation with all the authors. L.T. supervised the project.

# METHODS

## SAMPLES

<u>Nd-LSCO.</u> Single crystals of Nd-LSCO with a Sr content such that $p = 0.24$ were prepared with the floating zone technique at the University of Texas (by J.-S.Z.). A platelet sample was cut with dimensions $2 \times 0.5 \times 0.05$ mm$^3$ with the $c$ axis along the shortest dimension. Longitudinal contacts were made with silver epoxy annealed in oxygen for 1 hour at 500 C. The high symmetry crystallographic directions were determined with a precision better than 5° and they were normal to the faces of the sample. The superconducting critical temperature of this sample obtained from resistivity measurements in zero field, where $\rho = 0$, is $T_c = 10 \pm 1$ K.

<u>LSCO.</u> Single crystals of LSCO were grown with the floating zone technique, with a Sr content such that $p = 0.24$. Two samples were prepared, with similar dimensions and contacts to that of our Nd-LSCO sample. LSCO sample S1 was prepared by S.O. and sample S2 by H.T. Sample S1 was annealed for several weeks in oxygen flow to reduce oxygen deficiency, which yielded a lower $\rho_0$ compared to S2. The superconducting transition temperature of these two samples is $T_c = 17 \pm 1$ K and $16 \pm 1$ K for S1 and S2, respectively.

## TRANSPORT MEASUREMENTS

Electrical DC resistance was measured at Sherbrooke on all samples with an in-plane excitation current in the range of 0.5 to 2 mA and at steady field of 16 T applied normal to the CuO$_2$ planes.

The longitudinal resistance was measured with a conventional four-point configuration in pulsed fields up to 85 T in Toulouse. The in-plane excitation current was 5 mA or lower with a frequency range between 10 and ~60 kHz that was applied along the $a$ axis in all the samples.

A high-speed acquisition system was used to digitize the reference signal (current) and the voltage drop across the sample at a frequency of 500 kHz. The data were post-analyzed with software to perform the phase comparison [2].

## MAGNETORESISTANCE CALCULATIONS BASED ON BOLTZMANN MODEL

All the simulations are obtained by solving the Boltzmann equation below (all further details are discussed in refs. 3 and 22) :

$$\frac{1}{\rho_{xx}} = \frac{e^2}{4\pi^3} \oint d^2\boldsymbol{k}\, D(\boldsymbol{k})\, v_x[\boldsymbol{k}(t=0)] \int_{-\infty}^{0} v_x[\boldsymbol{k}(t)] e^{t/\tau} dt \,,$$

where the contour integral is over the Fermi surface, $D(\boldsymbol{k})$ is the density of state at point $\boldsymbol{k}$, $v_x$ is the component of the Fermi velocity in the x direction and the second integral is an integral of the Fermi velocity in the $x$ direction that calculates the probability that a quasiparticle with lifetime $\tau$ scatters after time $t$. The magnetic field enters through the Lorentz force and modifies the velocity by introducing a cyclotron motion to the quasiparticles.

**Extended Data Table 1** | The parameters that were used in the calculations.

| $t\ (meV)$ | $t'$ | $t''$ | $\mu$ | $p$ | $1/\tau_0$ $(ps^{-1})$ | $1/\tau_{aniso}$ $(ps^{-1})$ | $\nu$ | $\alpha$ |
|---|---|---|---|---|---|---|---|---|
| 160 | $-0.1364t$ | $0.0682t$ | $-0.8243t$ | 0.248 | 8.65 | 63.5 | 12 | 1.2 |

Tight-binding values that were taken from refs [3,20] to perform the calculations in this work: $t,\ t'$ and $t''$ are the first-, second- and third-nearest-neighbour hoping parameters, $\mu$ is the chemical potential, and $p$ is the doping. The remaining four parameters are from Eq. 1: $1/\tau_0$ is the amplitude of the isotropic scattering rate at $T = 0$, $1/\tau_{aniso}$ is the amplitude of the anisotropic scattering rate (which is $T$-independent), $\nu$ is the power of the cosine function, and $\alpha$ is a constant. The value of $1/\tau_0$ at $T > 0$ is: 9.45, 13.52, 15.10, 16.66, and 24.35 ps$^{-1}$ at $T = 4$, 30, 40, 50, and 100 K, respectively.

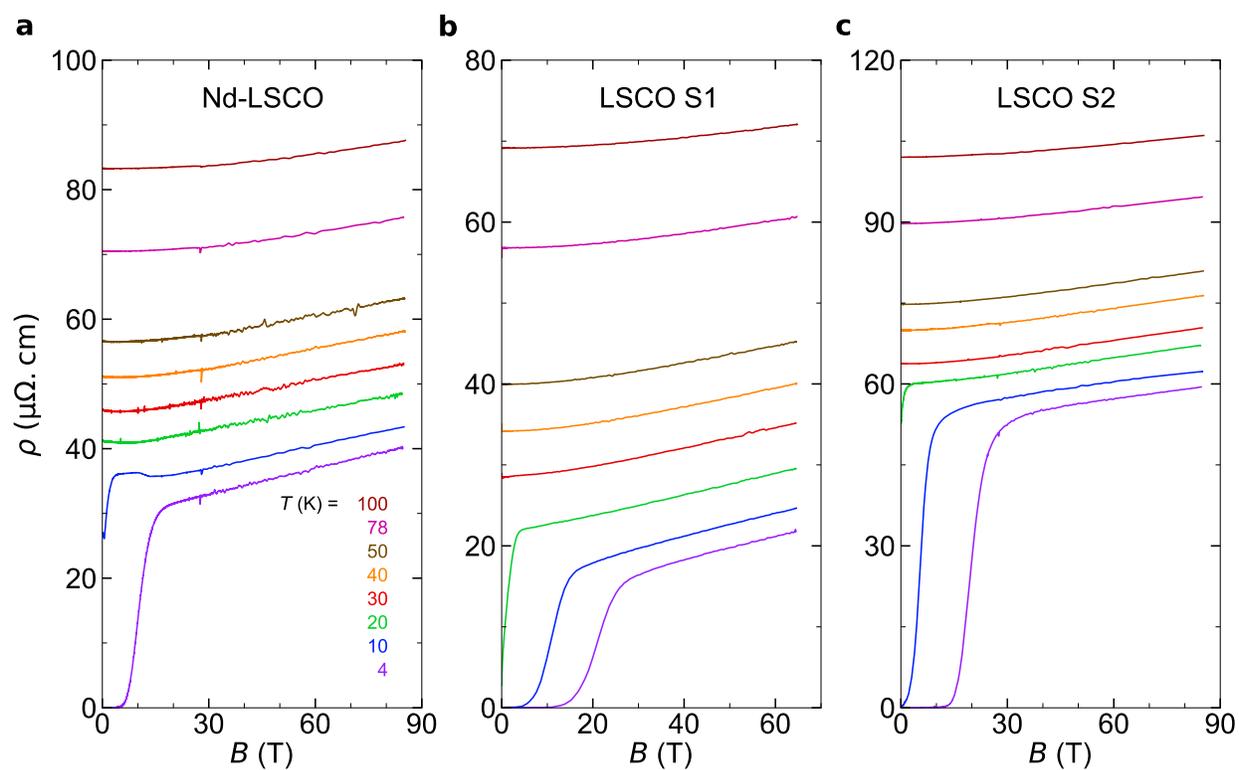

**Extended Data Figure 1 | Field dependence of resistivity.**

In-plane resistivity as a function of magnetic field ($B \parallel a$) in **a)** Nd-LSCO, **b)** LSCO sample S1 and **c)** LSCO sample S2, at the indicated temperatures.

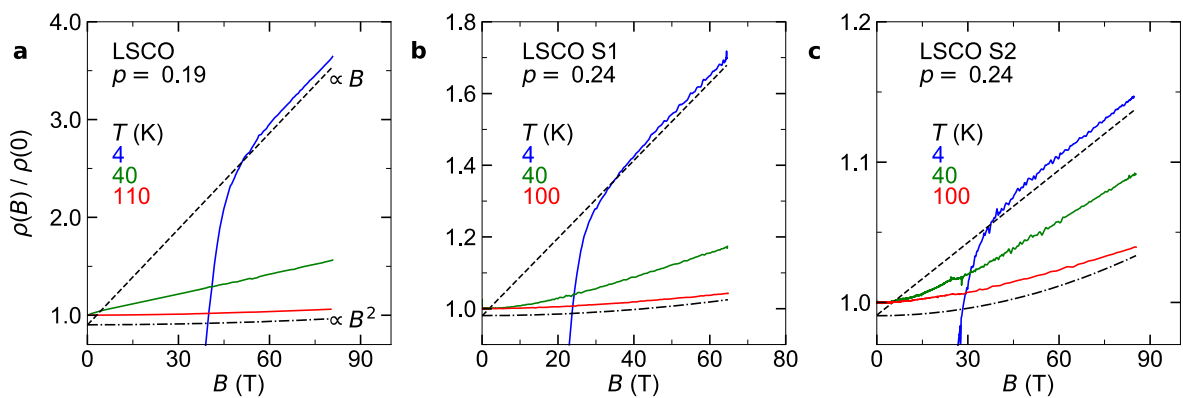

**Extended Data Figure 2 | Magnetoresistance in different LSCO samples.**

Magnetic field dependence of the in-plane resistivity ($J \parallel a$) in LSCO at $p = 0.19$ (**a**), from [4]) compared with **b)** LSCO sample S1 and **c)** LSCO sample S2 at $p = 0.24$, at temperatures as indicated. In all cases, $B \parallel c$.